\documentclass{elsart}
\usepackage{amssymb,epsfig,graphicx}
\begin{document}
\begin{frontmatter}

\title{Infra-red  Spectroscopic Studies of GdBaCo$_{2}$O$_{5.5}$}
\author{Shraddha Ganorkar},  \author{K. R. Priolkar\corauthref{krp}}\ead{krp@unigoa.ac.in}
\corauth[krp]{Corresponding author}
\address{Department of Physics, Goa University, Goa, 403 206 India.}

\begin{abstract}
This paper reports infrared spectroscopic studies on GdBaCo$ _{2} $O$ _{5.5} $ layered perovskite which exhibits successive magnetic transitions from paramagnetic to ferromagnetic to antiferromagnetic states as well as high temperature metal to insulator transition and a change in charge transport mechanism at low temperature. Infrared absorption spectra recorded at various temperatures in the range 80 K to 350 K reveal changes in the positions of Co-O stretching and bending frequencies which provide an explanation to the magnetic and transport behaviour of this compound.
\end{abstract}


\end{frontmatter}

\section{Introduction}
The study of transition-metal oxides has become a very attractive field during the last years.  They exhibit a
complex interplay between charge, spin, and lattice degrees of freedom which is the heart of many fascinating
phenomena like Colossal Magnetoresistance (CMR), superconductivity etc.  Among them, cobalt oxides are subject of
vivid studies because cobalt shows great facilities to present different spins states at different temperatures.

The spin-state degree of freedom of Co ions introduces new effects in these narrow band oxides\cite{jbg}.  The
existence of several possible spin states in given oxidation state makes cobaltites a rich but challenging system
to study. Co$^{3+}$ has three possible spin states: the low spin state (LS, t$_{2g}^{6}$ e$_{g}^{0}$, S=0), the
intermediate spin state (IS, t$_{2g}^{5}$ e$_{g}^{1}$, S=1) and high spin state (HS, t$_{2g}^{4}$ e$_{g}^{2}$,
S=2) which arise from the competition between crystal field, on-site Coulomb correlation and the inter atomic
exchange energy \cite{fro}.

The  RBaCo$ _{2} $O$ _{5.5} $ (R = rare earth) , in particular, displays a variety of interesting  phenomena
including metal insulator transition \cite{mar,mai}, spin state transition \cite{mor,res},  charge ordering
\cite{vog,sua} and giant magnetoresistance \cite{sua,tro}. The crystal structure consists of sandwiches CoO$_{2}$
- BaO - RO$_{0.5}$ - CoO$_{2}$ stacked along c-axis of orthorhombic lattice. The doubling of b-axis originates
from an alternation of CoO$_{5}$ square pyramids and CoO$_{6}$ octahedra along this direction. In contrast to
RCoO$_{3}$ \cite{pan,wudd,ber,pa}, the octahedra and pyramids are heavily distorted \cite{kha,pom}.  This
distortion supports a variety of Co$ ^{3+} $ spin state [LS, IS and HS] \cite{wu,zhi} as a function of
crystallographic environment and temperature.

A metal-insulator  transition (MIT) is present in several RBaCo$ _{2} $O$ _{5.5} $ compounds \cite{mai}. An
anomalous change of lattice parameters have been reported at the MIT temperature for GaBaCo$ _{2} $O$ _{5.5} $
and TbBaCo$ _{2} $O$ _{5.5} $ implying that the transition is related to changes in structure \cite{mor,fr}.
While magnetic susceptibility measurements indicate the MIT to be associated with spin state transition of
Co$^{3+}$ ion \cite{fro,mor,roy}. It has been proposed that the insulating state below MIT temperature is
promoted by the ordering of cobalt spin states with LS and IS states respectively in octahedrally and pyramidally
coordinated sites \cite{mar}. Further, using neutron diffraction studies it is suggested that, octahedrally
coordinated Co ions undergo a spin state transition from HS to IS or LS state in the insulating phase, while
pyramidally coordinated  Co ions remain in the IS state \cite{mor,fro}. Another scenario that has been put forth
based on electron diffraction and microscopy studies is the coexistence of IS (pyramids) and LS (octahedra) for T
$ < $  T$_{MI}$ (T$_{MI}$ - metal to insulator transition temperature) and HS Co$^{3+}$ at higher temperatures
\cite{mai}. Therefore, the spin state of Co ions on both sides of the MIT is still not clear.

Furthermore, several magnetic transitions have been reported for these compounds. Just below $T_{MI}$ the
compounds undergo a paramagnetic (PM) to  ferro(ferri)magnetic (FM) transition followed by a FM to
antiferromagnetic (AFM1) transition which is accompanied  by an onset of strong anisotropic magneto-resistive
effects and AFM1 to second antiferromagnetic (AMF2) phase transition. The mechanism of such magnetic
transformations at low temperature is still to be properly understood. It is also not clear why subtle changes in
oxygen content should cause drastic changes in magnetic properties \cite{kim,bur}.  Various contradicting
magnetic structure including different spin state of Co$^{3+}$ ions and also spin state ordering (SSO) have been
proposed, based on neutron diffraction \cite{pla,fau,soda,fron}and macroscopic measurements \cite{tas}. Muon-spin
relaxation ($\mu$SR) studies reveal that irrespective of the rare earth ion, a homogeneous FM phase with
ferrimagnetic SSO of IS and HS states develops through two first order phase transitions into phase separated
AFM1 and AFM2 phases with different types of antiferromagnetic SSO \cite{lue}. Density functional calculations
suggest that there is a strong hybridization between O-2p and Co-3d orbitals with a narrow charge transfer gap
near Fermi level, which gives rise to pd$\sigma$ hybridized hole in the O-2p valence band. With increasing
temperature, a gradual delocalization of the pd$\sigma$ holes in the almost HS  Co$ ^{3+} $ is responsible for
the successive magnetic transitions accompanied by spin reorientation in a spin-canted structure due to competing
FM and AFM interactions \cite{wu,wuh}. Resonant photoemission studies also support this picture \cite{fla}.

\section{Experimental}
The polycrystalline  sample of GdBaCo$ _{2} $O$ _{5.5} $ was prepared using conventional solid-state reaction
method. Gd$ _{2}$O$ _{3} $ and BaCO$  _{3}$ were preheated at 900$^\circ$C and 700$^\circ$ C respectively for 12
hours. Stoichiometric amount of Gd$  _{2}$O$ _{3} $, BaCO$  _{3}$ and Co(NO$ _{3} $)$ _{3} $.6H$ _{2} $O were
dissolved in  HNO$  _{3}$. This mixture was stirred continuously using a magnetic stirrer for about 30 minutes to
ensure proper mixing. The solution was allowed to settle for an hour and dried on a heater. This dried powder was
then sintered in  a furnace at 1000$^\circ $ C for 12 hours twice with an intermediate grinding to ensure
homogeneity. Finally the sintered powder was ground again and pressed into pellets and annealed in oxygen
atmosphere at 1050$^\circ$ C for 24 hours. The sample was furnace cooled at a rate of 1$ ^\circ$C per minute.
The sample was deemed to be phase pure, as X-ray diffraction (XRD) data collected on a Rigaku  D-Max IIC X-ray
diffractometer  in the  range of $ 20^\circ \le 2\theta \le 80^\circ$ using CuK$\alpha$ radiation showed no
impurity reflections. The diffraction pattern was Rietveld refined using FULLPROF suite and structural parameters
were obtained. The oxygen content verified by iodometric titration, was found to be $ \delta $=5.5. The
magnetization measurements were carried out as a function of temperature and magnetic field using a Quantum
Design SQUID magnetometer at fields of 1000 Oe, in the temperature range of 5 K to 300 K. The magnetization
measurements were carried out in both the zero-field cooled (ZFC) and field-cooled modes(FC). The temperature
(100 K to 300 K) dependence of the electrical resistivity was measured using a standard four probe set up and a
Keithley 2182 nanovoltmeter and Keithley constant current source. IR spectra were recorded on FTIR-8900 Shimadzu
spectrophotometer with Oxford OPTIDNVA Cryostat with KRS5 windows in the temperature range 80 K to 350 K. These
KRS5 windows give a sharp peak like feature around 675 cm$^{-1}$.

\section{Results and Discussion}
\begin{figure}
\includegraphics[scale=1.5]{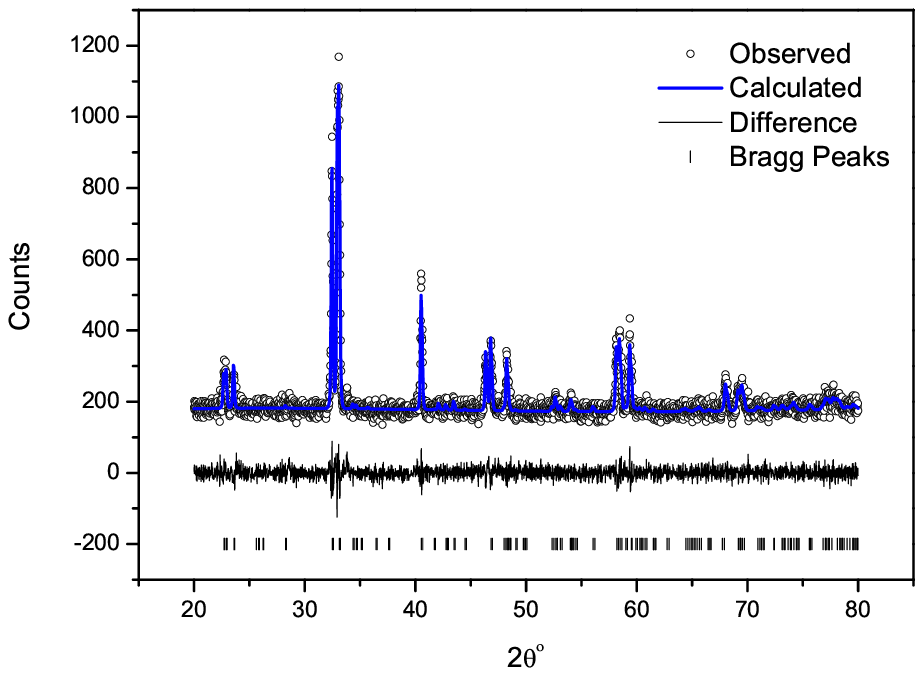}
\caption{\label{xrd}Rietveld fitted XRD pattern of GdBaCo$_2$O$_{5.5}$. Circles represent experimental data, continuous line through the data is the fitted curve and the difference pattern is shown at the bottom as solid line.}
\end{figure}

XRD pattern presented in Fig. \ref{xrd} shows the formation of a single phase sample.  Rietveld refinement of the
XRD pattern indicates that structure is orthorhombic with Pmmm space group. The lattice parameters obtained from
fitting are $a$  = 3.8760(4)\AA, $b$ = 7.822(1)\AA~ and $c$ 7.533(1)\AA which agree well with those reported in
literature for GdBaCo$_2$O$_{5.45}$ \cite{roy}.  Iodometric titration reveals the value of  $\delta$ = 0.5.

\begin{figure}
\includegraphics[scale=1.5]{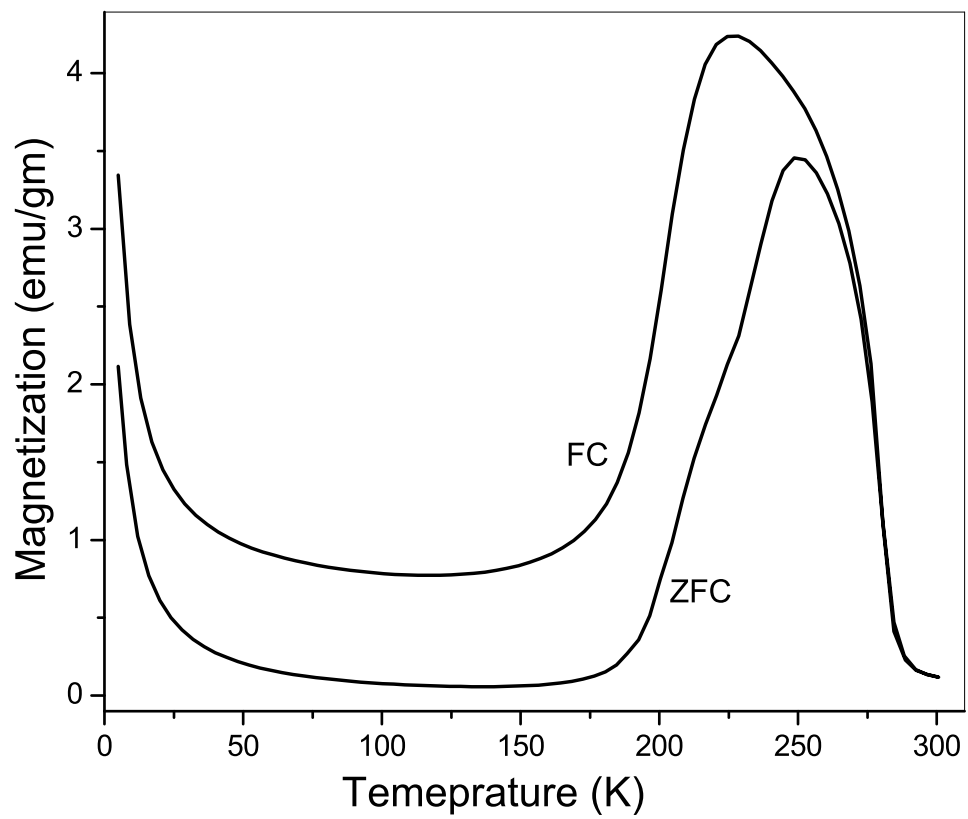}
\caption{\label{Mag}Plot of magnetization versus temperature for GdBaCo$ _{2} $O$ _{5.5} $.}
\end{figure}

The temperature dependence of magnetization M(T) is shown in Fig.\ref{Mag}.  Upon cooling M(T) in both ZFC and FC
cycles increases rapidly indicative of a PM to FM transition at T$ _{c} $ = 275 K. It reaches a maximum at around
250 K and starts decreasing with further lowering of temperature. Around T $\simeq$ 230 K  and T= 210 K there are
appearances of two magnetic transitions which are reported to be AFM. At still lower temperatures (T $ \leq $ 100
K) the magnetization shows paramagnetic behaviour which is attributed to Curie-Weiss behaviour of Gd sub-lattice
\cite{roy}.

\begin{figure}
\includegraphics[scale=1.5]{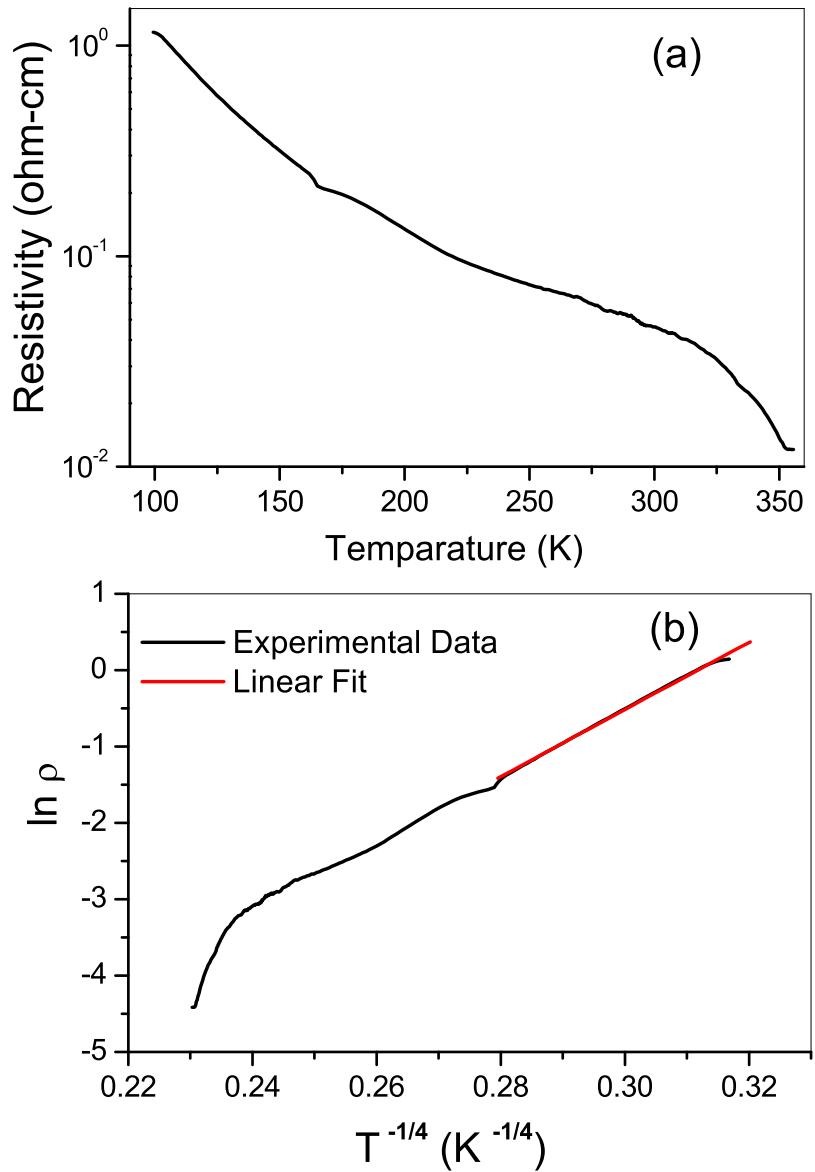}
\caption{\label{Res}(a) Plot of resistivity as a function temperature for GdBaCo$ _{2} $O$ _{5.5} $ (b) Resistivity versus T$^{-1/4}$ data for GdBaCo$ _{2} $O$ _{5.5} $ fitting to VRH model.}
\end{figure}

The resistivity as a function of temperature is plotted in Fig.\ref{Res}(a).  At 350 K there is a clear signature
of MI transition followed by a broad hump which coincides with PM - FM transition. Further, one more anomaly is
seen at 164 K which has been attributed to strong coupling between the spin order and charge carriers \cite{zho}.
Below T $  < $ 164 K resistivity follows Mott's  variable range hopping (VRH) model  as can be clearly seen in
Fig \ref{Res}(b).

\begin{figure}
\includegraphics[scale=1.5]{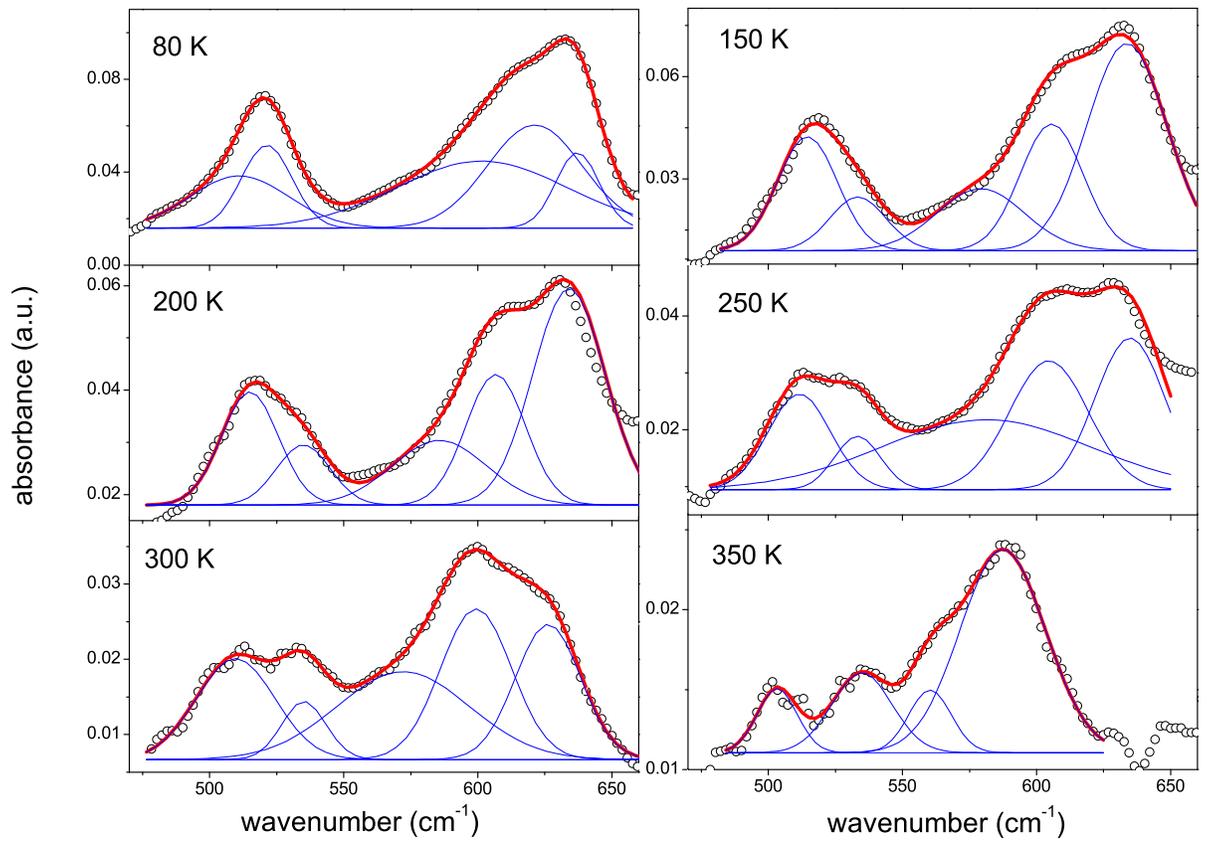}
\caption{\label{IR}IR absorption spectra recorded at various temperatures for GdBaCo$ _{2} $O$ _{5.5} $ along with best fitted curves (solid line) which is a sum of four or five constituent Gaussian peaks.}
\end{figure}

Fourier transform infrared (IR) absorption spectra at various temperatures in the range 350 K $\le$ T $\le$ 80 K
are presented in Fig.\ref{IR}. The optical phonon bands appear in the range 500 cm$^{-1} $ to 600 cm$ ^{-1} $.
RBaCo$_{2}$O$_{5.5}$ compounds being derivatives of RCoO$_{3}$ single perovskites, the same nomenclature can be
used to classify the absorption modes. As per the nomenclature available for RCoO$_3$ compounds, there are eight
IR active modes reported for RCoO$ _{3} $ system of which  the modes around 600 cm$ ^{-1} $ are called stretching
modes and those around 500 cm$ ^{-1} $ are termed as bending modes. The stretching modes can be fitted with three
Gaussian's and one Lorentzian \cite{sud}.

In RCoO$_3$, Co$^{3+}$ is in LS state at low temperatures.  With increase in temperature, the crystal-field
splitting decreases and two excited states (IS and HS )can be populated \cite{ima,nek}. Since these spin states
are very close in energy, transitions to these higher spin state can occur either with variation in temperature
or with deformation of the crystal lattice. The frequencies of IR vibration modes exhibit anomalies at these spin
state transition temperatures. Generally the vibration modes harden when the sample undergoes a transition from
IS/HS to LS state. If the magnetic transitions observed in GdBaCo$_2$O$_{5.5}$ are due to spin state
transformation then its lattice vibration modes should also show similar temperature dependence.

\begin{figure}
\includegraphics[scale=1.5]{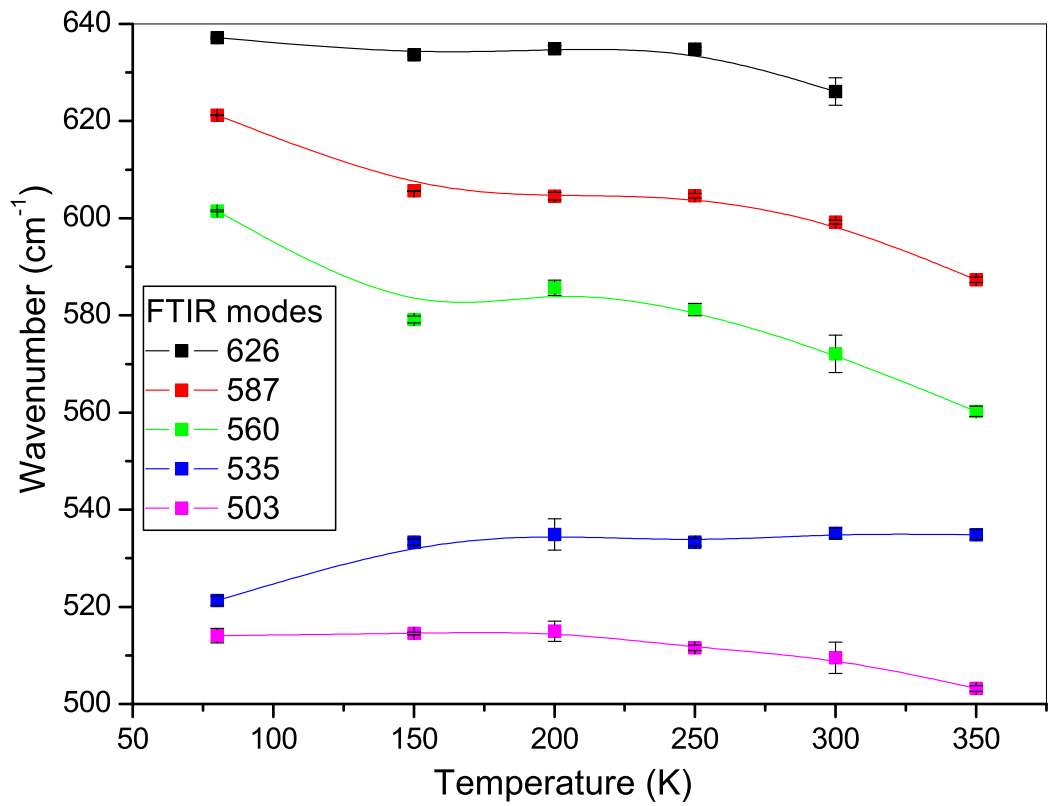}
\caption{\label{IRmodes} Evolution of  IR stretching modes with temperature for GdBaCo$ _{2} $O$ _{5.5} $.}
\end{figure}

In order to examine the temperature dependence of frequencies of Co-O stretching and bending modes in
GdBaCo$_2$O$_{5.5}$, the IR absorption spectra from 500 to 650 cm$^{-1}$ were fitted with Gaussian peaks. At 350
K the IR spectra can be fitted with  four Gaussians, bending modes centred at 503, 535 cm$^{-1}$ and stretching
modes at 560 and 587 cm$^{-1}$ . At lower temperature a new stretching mode appears at 626 cm$^{-1}$ due to anomalous cell contraction
as a consequence of the metal to insulator transition \cite{fron}. The observed absorption modes are in agreement with the IR absorption bands reported for TbBaCo$
_{2} $O$ _{5.5}$ \cite{kas}. The spectra at 300 K and lower temperatures can be best fitted with five
Gaussian peaks at 510, 534, 571, 600 and 626 cm$^{-1}$. The fitting to the experimental data is shown as
continuous line in Fig. \ref{IR}.  The temperature evolution of the IR stretching frequencies is plotted in
Fig.\ref{IRmodes}. The bending mode at 503 cm$^{-1}$ shifts continuously toward higher wavenumber from 350K to
150 K and then shows a decrease in its frequency below 150 K. The second bending mode at 534 cm$ ^{-1} $ has
nearly temperature independent evolution down to 150K, below which it shifts towards lower frequencies and tends to
merge with the other bending mode.  The stretching modes 560, 588 and 626 cm$^{-1}$ at 350 K show a gradual increase in wavenumber. At 150 K the stretching  modes show shift towards lower wavenumber, while at 80 K these modes harden again. 

At 350 K the compound is a PM metal with Co$ ^{3+}$ in HS state. As temperature decreases it undergoes a MI
transition followed by PM to FM transition. These transitions are believed to be due to a change in Co$^{3+}$
spin state from HS to IS concomitant with decrease in cell volume. These changes are reflected in IR absorption
spectra by an increase in vibration frequency of 560 and 588 cm$^{-1}$ stretching modes by 12 cm$^{-1}$ and
occurrence of a new mode at 626 cm$ ^{-1}$. Very similar results have been reported in case of RCoO$ _{3}$
compound where it is  observed that the IR vibration modes harden with decrease in temperature
\cite{sud,mort,taj}.

It can be seen that the stretching modes (560, 588 and 626 cm$^{-1}$) shift towards lower wavenumber below 200 K which
coincides with the appearance of AFM order and change in resistivity behaviour respectively.  This  softening of
mode could be related to a larger deformation of Co-O polyhedra resulting in displacement of Co ions away from
the ideal middle plane. This results in higher bending of Co-O-Co bonds.  Reduction in Co-O-Co bond angle
localizes the p-d holes favouring antiferromagnetic order. Such a change in Co $3d$ and O $2p$ hybridization has
been predicted by density functional theory calculations as well observed in resonant photoemission studies
\cite{wu,wuh,fla}. The reduction of Co-O-Co bond angle will reduce the electron transfer integral through Co-O-Co
network leading to higher resistivity. Indeed as can be seen from Fig. \ref{Res}, resistivity below 164 K rises
steeply and it follows the Mott's VRH law. 

\section{Conclusion}
In  conclusion, IR spectra of  GdBaCo$ _{2} $O$ _{5.5} $ shows changes in positions and intensities of Co-O stretching and bending which can be related to various magnetic transitions. Upon MI followed by PM to FM transition a stretching mode at appears at 626 cm$^{-1}$ due to Co$^{3+}$ HS $\to$ IS spin state transition and a decrease in cell volume. Further, IR absorption studies indicate that AFM transition is due to increase in p-d hybridization. This increase of p-d hybridization can also explain the change in resistivity behaviour at low temperatures.

\newpage

\end{document}